\documentclass[doublecol]{epl2} % for 2 columns style with line numbers
\def\be{\begin{equation}}
\def\ee{\end{equation}}

\newlength{\textlarg}

\usepackage{ulem}
\renewcommand{\etal}{{\it et al.}}

\author{Daniel Suchet \inst{1} \and Mihail Rabinovic  \inst{1} \and Thomas Reimann  \inst{1}  \and Norman Kretschmar  \inst{1}  \and Franz Sievers  \inst{1}  \and Christophe Salomon \inst{1}  \and Johnathan Lau \inst{2} \and Olga Goulko  \inst{3}  \and Carlos Lobo  \inst{2}  \and Fr\'ed\'eric Chevy  \inst{1}}
\shortauthor{Suchet \etal}

\institute{
  \inst{1} Laboratoire Kastler Brossel, ENS-PSL Research University, CNRS, UPMC, Coll\`ege de France, 24, rue Lhomond, 75005 Paris\\
  \inst{2} Mathematical Sciences, University of Southampton, Highfield Southampton, SO17 1BJ,UK\\
  \inst{3} Department of Physics, University of Massachusetts, Amherst, MA 01003, USA
}

\title{Analog Simulation of Weyl Particles with Cold Atoms}


\date{\today}
\abstract{
In this Letter we report on a novel approach to study the dynamics of harmonically confined Weyl particles using magnetically trapped fermionic atoms. We find that after a kick of its center of mass, the system relaxes towards a steady state even in the absence of interactions, in stark contrast with massive particles which would oscillate without damping. Remarkably, the equilibrium distribution is non-Boltzmann, exhibiting a strong anisotropy which we study both numerically and experimentally. }

\begin{document}

\maketitle

\section{Introduction} Weyl fermions were introduced for the first time in 1929 as massless solutions of the Dirac equation \cite{Weyl1928Elektron}. Despite constituting one of the paradigms of contemporary high energy physics, their existence in nature has remained unconfirmed until very recently. While at first suggested to describe neutrinos, the observation of flavor oscillations implying a non-zero rest mass ruled out this hypothesis \cite{barger2012physics}. It had been pointed out that they could be observed in the form of low energy excitations of crystalline structures with a linear dispersion relation around a so-called Weyl point. The non-trivial topology of such Weyl semimetals is responsible for the Adler-Bell-Jackiw chiral anomaly \cite{Adler1969,Bell1969} which leads to remarkable properties such as negative magnetoresistance, anomalous Hall effect and non-local transport \cite{Hosur2013}. Moreover, the confinement of quasiparticles obeying a linear dispersion relation was suggested as a way to engineer individual quantum dots \cite{silvestrov2007quantum}, notably for the improvement of multiple exciton generation in solar cells \cite{Delerue2011}.

The mere existence of Weyl points in reciprocal space requires a broken time-reversal or inversion symmetry, which are challenging to implement experimentally. As a consequence, observations of Weyl particles were reported only recently in 3D-compounds such as \mbox{HgCdTe}, \mbox{HgMnTe} \cite{Orlita2014}, TaAs \cite{xu2015discovery, Huang2015} as well as in photonic crystals \cite{Lu15experimental}.
Owing to their high degree of control and versatility, cold atoms offer a promising and complementary route for the experimental study of Weyl fermions. Early proposals in this context were based on the band structure of cold atoms in 3D optical lattices extending the 2D Harper Hamiltonian \cite{Dubcek2015}. Yet another approach is analog simulation where one takes advantage of the mathematical equivalence between two seemingly different physical systems. Such mapping were successfully used in the past to relate, for instance, Anderson localization to the $\delta$-kicked rotor \cite{grempel1984quantum,casati1989Anderson,Garreau2008}, quantum magnetism to the filling factor of an optical lattice \cite{sachdev02mott,simon2011quantum}, the solutions of Dirac equation to the dynamics of ion chains   \cite{gerritsma2010quantum,Gerritsma2011}, or quantum Hall edge states to the eigenmodes of classical coupled pendula \cite{susstrunk47observation}.

In this letter, we report on the analog simulation of Weyl particles in a harmonic potential using a dilute gas of cold magnetically trapped atoms. Using a canonical mapping exchanging position and momentum in the system's Hamiltonian, we address the dynamics of an ensemble of non-interacting Weyl particles after excitation of their center of mass.  The system's ensuing relaxation towards a steady-state exhibits intriguing dynamics, resulting in a strongly anisotropic and non-thermal momentum distribution of the cold gas. Our observations are interpreted using a kinetic model based on virial theorem and energy conservation.

\section{Mapping}
The magnetic quadrupole trap is a common technique for confining neutral atoms \cite{migdall1985first}. It is made up of a pair of coils carrying anti-parallel currents, creating close to their symmetry center a linear magnetic field ${\bf B_0(r)} = b(\alpha_x x,\, \alpha_y y,\, \alpha_z z)$, where $z$ is the symmetry axis of the coils. Here $b$ denotes the magnetic field gradient and Maxwell's equations imply that $\alpha_x=\alpha_y=1$, $\alpha_z=-2$. For a spin $1/2$ atom of mass $m$ carrying a magnetic moment $\mu$, the coupling to this field leads to the  single-particle Hamiltonian
\be
  h \left( \bm{ r},\bm{ p} \right) 	=	\frac{p^2}{2m}-\mu \,\bm\sigma\cdot \bm B_0(\bm r), \label{eq:H0}
\ee
where  $\bm\sigma$ are the Pauli matrices. By means of the canonical mapping $X_i = cp_i/\mu b\alpha_i \,$ and $P_i=-\mu b\alpha_ix_i/c$ with $c$ being an arbitrary velocity scale, the Hamiltonian (\ref{eq:H0}) becomes
\be
	H = c \,\bm\sigma\cdot\bm P + \frac{1}{2} \sum_i k_i X_i^2. \label{Eq:H0Weyl}
\ee
The first term corresponds to the  kinetic energy $c\bm\sigma\cdot\bm P$ of a massless Weyl particle moving at velocity $c$ while the second one is readily identified as an anisotropic harmonic potential, characterized by spring constants $k_i=\alpha_i^2\mu^2 b^2/mc^2=\alpha_i^2 k$ along each direction $i$. This mapping is at the core of our work and it shows that neutral atoms confined by a linear potential can be used to simulate experimentally the dynamics of Weyl particles.

The single-particle trajectories of the Weyl  particles can be obtained using Ehrenfest's theorem applied to the Hamiltonian (\ref{Eq:H0Weyl}). Using uppercase (lowercase) symbols for the phase-space coordinates of the Weyl particles (spin-1/2 atoms), we obtain respectively in the Heisenberg representation:

\noindent\begin{minipage}{0.49\columnwidth}
	\begin{eqnarray}
	\dot X_i&=&c\sigma_i\label{Eq:velocity}\\
	\dot P_i&=&-k_i X_i\\
	\dot{\bm\sigma}&=&\frac{2c}{\hbar}\bm\sigma\times\bm P \label{Eq:precession}
	\end{eqnarray}
\end{minipage}
\hfill
\begin{minipage}{0.49\columnwidth}
	\begin{eqnarray}
	\dot p_i&=&\mu b \alpha_i\sigma_i\\
	\dot x_i&=& p_i/m\\
	\dot{\bm\sigma}&=&\frac{2\mu}{\hbar} \bm\sigma\times\bm B(\bm r) \label{Eq:precession2}
	\end{eqnarray}
\end{minipage}
\bigskip

Equations (\ref{Eq:velocity}) to (\ref{Eq:precession2}) are fully quantum, but in the following we will focus on the classical regime, and consider the operator mean values. Noting that $\langle \bm\sigma\rangle^2=1$, Equation (\ref{Eq:velocity}) immediately shows that even in a harmonic trap Weyl particles move at a constant velocity $c$. Equations (\ref{Eq:precession}) and (\ref{Eq:precession2}) describe respectively the particle's spin precession around the momentum $\bm P$ and magnetic field $\bm B$. The adiabatic following results in the conservation of helicity and of the Zeeman populations, giving rise to topological properties. The analogy existing between these two equations allows to draw a parallel between a peculiar feature of Weyl particles, the Klein paradox \cite{Klein1929}, and the well known Majorana losses \cite{Majorana1932, Bergeman1989, Petrich1995,sukumar1997} for magnetic traps. The Klein paradox states if the rate of change of the particle's energy is too high (i.e. much larger than $2Pc/\hbar$ for Weyl particles), the spin will not follow the momentum adiabatically and the helicity of the particle is not conserved. The resulting transfer of the particle to negative energy states leads to dramatic effects, such as the suppression of back-scattering for electrons in 1D carbon nanotubes \cite{ando1998berry}. For the equivalent picture of magnetically trapped atoms, in regions where the Larmor frequency $2 \mu B/\hbar$ is smaller than the rate of change of the Zeeman energy, the atomic spin will not follow adiabatically the direction of the local magnetic field. This results in Majorana losses. The absence of backscattering in carbon nanotubes then appears as equivalent to the impossibility to trap atoms in a 1D magnetic quadrupole. Furthermore, for an ensemble of particles at temperature $T$, we can define a Klein loss rate $\Gamma_{\rm Klein}$ equivalent to the Majorana rate $\Gamma_{\rm Maj.}$:
	\be
		\Gamma_{\rm Maj.} \simeq \frac{\hbar}{m} \left(\frac{\mu_B b}{k_B T}\right)^2\quad,\quad
		\Gamma_{\rm Klein} \simeq \hbar k \left(\frac{c}{k_B T}\right)^2
	\ee
Just like Majorana losses prevent the existence of a true thermodynamic equilibrium in a quadrupole trap, the Klein paradox prevents stable trapping of Weyl particles in external potentials \cite{silvestrov2007quantum}. Nevertheless, at high enough temperature such as considered in our experiments below, particles spend little time close to 0 and we can neglect Majorana-Klein losses. Particles of positive and negative helicities can therefore be described by the effective Hamiltonians:
\begin{equation}
	H_\pm = \pm c|P| + \sum_i \frac{k_i X_i^2}{2}. \label{eq:H0'}
\end{equation}
The negative-helicity Hamiltonian $H_-$ is not bounded from below which implies diverging trajectories. This directly corresponds to the anti-trapped high-field seeking states of the atomic problem. In the following we shall therefore restrict our study to the case of metastable, positive-helicity particles.

\section{Results}
Using the mapping derived above, we explore the dynamics of Weyl particles using a sample of  spin-polarized $^6$Li atoms confined in a quadrupole magnetic trap.

The experimental preparation of the sample starts with a dual species magneto-optical trap which is loaded with fermionic $^{6} \mathrm{Li}$ and $^{40} \mathrm{K}$. In a second step the clouds are subjected to blue detuned D1 molasses \cite{fernandes2012sub,sievers2015simultaneous}, cooling both species down to the $50 \mu K$ regime. Subsequently the atoms are optically pumped into their low-field seeking stretched Zeeman states $\left|F=3/2,m_F=3/2\right\rangle$ and $\left|9/2,9/2\right\rangle$, respectively. Finally, we ramp a magnetic quadrupole field up to $b = 80 \, \mathrm{G/cm}$ within 500\,ms, capturing $10^7$ $^{6} \mathrm{Li}$ and $10^9$ $^{40} \mathrm{K}$ atoms. Inter-species- as well as p-wave collisions among $^{40} \mathrm{K}$ atoms \cite{demarco1999measurement} allow for the complete thermalization of the two clouds at approximately $T_0 = 300 \, \mathrm{\mu K}$. This value is high enough to preclude Majorana losses during the experiment's duration and is well below the p-wave collision threshold. %This is a convenient working temperature since lower values would imply noticeable Majorana losses within the Li sample.
After thermalization the $^{40} \mathrm{K}$ atoms are removed from the trap by shining in resonant light, which leaves $^{6} \mathrm{Li}$ unaffected.

%The  field gradient is $b=80$~G/cm \cite{ridinger2011large} and the initial temperature of the cloud is $T_0=300\,\mu$K \cite{sievers2015simultaneous}.   Atoms are spin-polarized in the hyperfine state $|F=3/2,m_F=3/2\rangle$. Therefore the cold gas can be considered as an ensemble of non interacting particles. While, strictly speaking, these atoms are not spin 1/2 particles, they are however described by the same Hamiltonian $H_+$ as positive helicity Weyl particles. They will consequently exhibit the same dynamics with $\mu=\mu_B$, where $\mu_B$ is the Bohr Magneton.

\begin{figure}\centering
	\includegraphics[width=0.80\columnwidth]{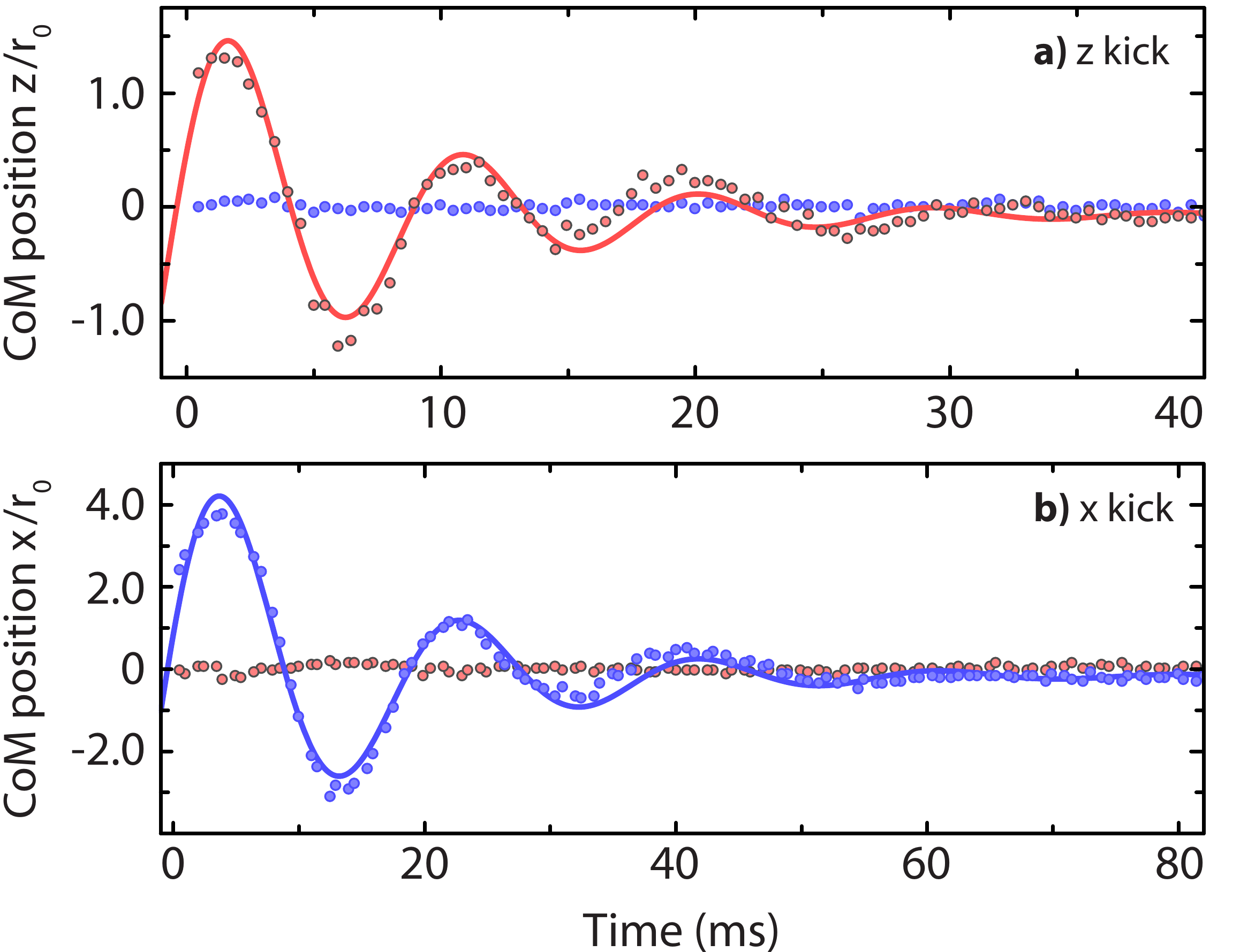}
	\caption{Center-of-mass oscillations of the Lithium cloud after a kick along the symmetry axis of the coils $\bm{z}$ (a) and along $\bm{x}$, within the symmetry plane. (b). Blue squares (resp. red circles) are experimental data along $\bm{x}$ (resp. $\bm{z}$). Solid lines are exponentially damped sinusoidal oscillations. Damping results solely from dephasing of single particle trajectories. Equivalently, this corresponds to momentum oscillations of Weyl particles in a harmonic trap. Here, $r_0=k_B T/\mu_B b\sim 0.6$ mm  and $t_0 = \sqrt{mk_B T}/\mu_B b\sim1$ ms.}
	\label{fig:exposc}	
\end{figure}

We deliver a momentum kick to the cloud by quickly turning on a magnetic bias field $\bm{B}$ which shifts the center of the trapping potential by a distance ${\boldsymbol{\delta}}$ for a short time $\tau$. Maximum trap center displacements are of order $\delta\sim 7\, r_0$ along $\bm x$ and $\delta\sim 5\, r_0$ along $\bm z$, where $r_0=k_B T/\mu_B b\sim0.6$ mm is the characteristic thermal size of the cloud. The kick duration $\tau$ is typically a few ms, being constrained by the coil inductances and eddy currents in the surrounding vacuum chamber. During the kick, the ensemble acquires an overall momentum of magnitude $q \sim \mu_{\mathrm{B}} b \tau$, similar to free fall in gravity. The potential is then quickly brought back to its initial position, and   the cloud is left to evolve during a variable time $t$ before switching off all fields to perform a time of flight measurement of the momentum distribution. Temperatures and kick velocities are measured with a time-of-flight (TOF) technique: the trapping potential is abruptly switched-off and the atomic cloud expands freely during a few ms, before it gets imaged on a CCD camera by resonant light absorption. The center of mass velocity can be extracted by tracking the center of the distribution during the TOF, while the temperature is measured using the standard deviation of the position distribution for sufficiently long TOF expansion times.
A limitation for the accurate determination of the kick amplitude originates from transient currents lasting about 3\,ms, which appear while abruptly switching off the quadrupole magnetic trap with gradients of the order of $100\,$ G/cm. The transient magnetic field creates a position-dependent Zeeman effect which deforms the atomic cloud profile at short TOF durations.
This results in a potential error in the measurement of the center of mass momentum with or without kick. For instance,  in the absence of a kick we observe a small parasitic velocity $v_0$ which is proportional to the magnetic gradient $b$ and reaches $30\,$cm/s at our highest value $b= 165\,$G/cm. Therefore, to infer the actual momentum delivered  to the cloud solely by the kick, we subtract $v_0$ measured after the thermalization time of $500\,$ms from the velocity right after the kick. The fit errors are given by the error bars in Fig. 2 and account for our statistical errors of typically $0.05 /m k_\mathrm{B}$ on temperature. Performing the experiment with 4 different magnetic field gradients, we estimate a systematic uncertainty of $0.2 / m k_\mathrm{B}$ for the fitted coefficient of the parabolic dependence of the heating on the momentum kick strength in Fig. 2.

For Weyl fermions, this excitation corresponds to displacing the Weyl point in momentum space, waiting until the distribution has moved by a distance $\bm{R}$ and switching the Weyl point back to its initial position. The resulting time evolution of the position (resp. momentum) distribution of the Lithium atoms (resp. Weyl particles) is shown in Fig.1. Even though collisions are absent, oscillations are damped as a consequence of the dephasing between single particle trajectories. The initially imparted energy is converted into internal energy of the cloud and the distribution reaches a steady state within a few units of time $t_0 = \sqrt{mk_B T}/\mu_B b\sim1$ ms.

To characterize the steady state, we kicked the cloud along the $\bm{z}$- and $\bm{x}$-directions and measured (i) the center of mass velocity right after the kick and (ii) the respective steady state momentum distribution after a sufficiently long relaxation time, typically 250 $t_0$. We define the steady-state's effective temperature along direction $i$ as the second moment of the momentum (resp. position) distribution:
\begin{equation}
k_B T_i=\frac{\langle p_i^2\rangle}{m}=k_i\langle X_i^2\rangle,
\end{equation}
where $\langle\cdot\rangle$ denotes the statistical average.

The heating $\Delta T$ and the center of mass momentum $\bm{q}$ induced by the momentum kick are extracted from the difference between the corresponding values at quasi-equilibrium and the ones measured right after the kick. While for a fully thermalized system the temperatures in both directions should be equal, our results presented in Fig.2, show a very strong anisotropy, thus demonstrating that the final distribution is non-thermal. The temperature increases much more in the direction of the kick than in the transverse directions. A kick in the $\bm{z}$ direction produces strong heating along $\bm{z}$, but a much weaker energy transfer along $\bm{x}$. Conversely, a kick in the $(\bm{x},\bm{y})$ plane results in smaller heating in the $\bm{z}$ direction than along $\bm{x}$. Quantifying the strength of the kick through the dimensionless parameter
\begin{equation}
\eta=\frac{\langle q\rangle }{\sqrt{mk_B T_0}}=\sqrt{\frac{\sum_i k_i \langle R_i\rangle^2}{k_B T_0}},
\end{equation}
we find that for kicks along $\bm{x}$ the best quadratic fits are given by $\Delta T_\mathrm{x} / T_0 = 0.52 (5)_\mathrm{stat} (20)_\mathrm{syst} \times \eta^2$ and $\Delta T_\mathrm{z} / T_0 = 0.10 (4)_\mathrm{stat} (5)_\mathrm{syst} \times \eta^2$. For kicks along the strong axis $\bm{z}$, $\Delta T_\mathrm{z}/T_0 = 0.63 (7)_\mathrm{stat} (20)_\mathrm{syst} \times \eta^2$ and $\Delta T_\mathrm{x} /T_0 = -0.14 (5)_\mathrm{stat}(8)_\mathrm{syst} \times \eta^2$.

\begin{figure}\centering
	\includegraphics[width=1\columnwidth]{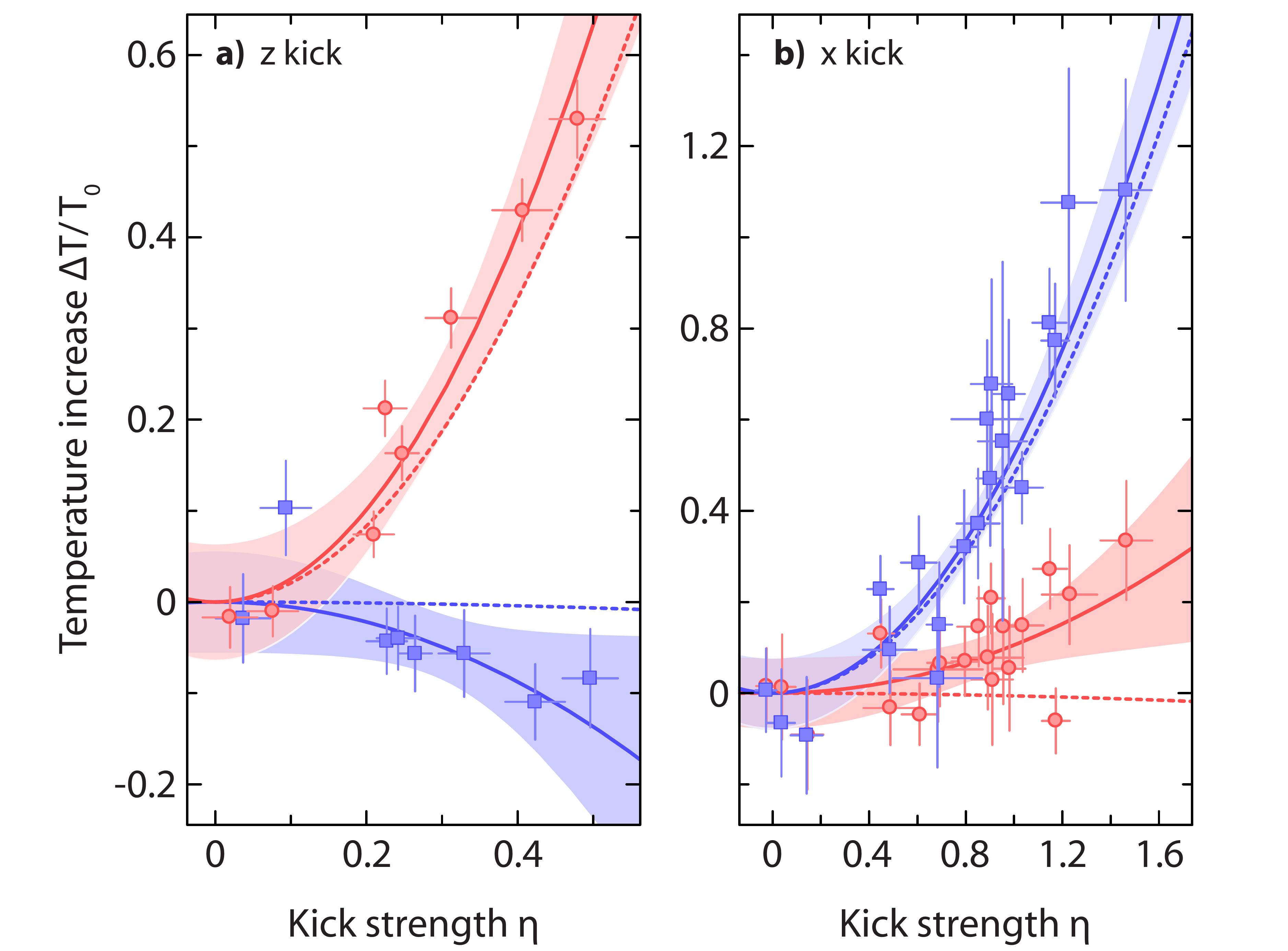}
	\caption{Temperature increase $\Delta T/T_0 = (T - T_0)T_0$ along $\bm{x}$ (blue squares) and $\bm{z}$ (red circles) as a function of the normalized CoM momentum $\eta$ acquired during the kick. \textbf{a)} $\bm{z}$ momentum kick at $b=70\,\mathrm{G/cm}$. \textbf{b)} $\bm{x}$ momentum kick at $b=55\,\mathrm{G/cm}$. Solid lines are quadratic fits to the experimental data with coefficients given in the text. Error bars represent the temperature statistical uncertainty and shaded zones give the 95\% confidence level of the fits. Dashed lines are results of numerical simulations presented in figure \ref{fig:simukick}.}
	\label{fig:expresults}	
\end{figure}

\subsection{Numerics}
In order to interpret these results, we performed single particle dynamics simulations on an ensemble of $10^5$ particles. As in the experiment, an excitation is applied to the initial distribution by displacing the trap center (resp. the Weyl point in momentum space) by an amount $\delta$ for a duration $\tau$ before bringing it back to its initial position. To simulate the effect of coil response time and eddy currents, we consider excitations of constant duration and increasing displacement. The simulation does not include any collisions, and yet we observe, as in the experiment, a relaxation towards a steady state after $\sim 100 \, t_0$ as all calculated moments of the distribution up to 8$^{\rm th}$ order reach a stationnary value. We also reproduce the strong anisotropy between the $\bm{z}$ and $\bm{x}$ direction (see Fig. 3). Numerical simulations also provide access to the $\bm{y}$ direction (not measured in the experiment), which also appears to be decoupled from the strong axis $\bm{z}$, but reaches the same final effective temperature as the other weak axis $\bm{x}$, regardless of the kick direction. The simulated dynamics thus features a quasi-thermalization within the symmetry plane of the distribution.

\begin{figure}\centering
	\includegraphics[width=1\columnwidth]{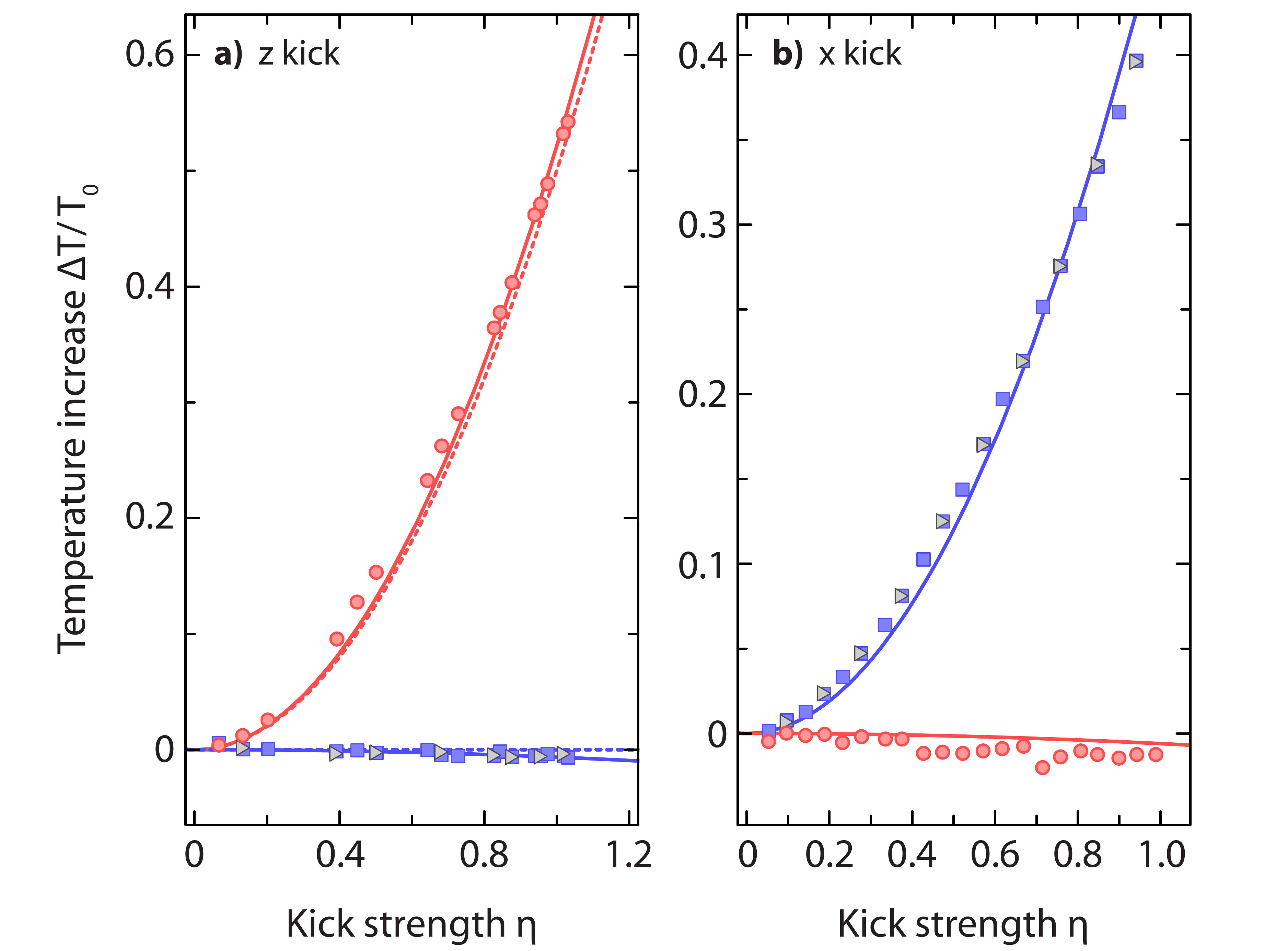}
	\caption{Numerical simulation of the temperature increase as a function of the normalized CoM momentum kick $\eta$. Data points are obtained by solving the classical equations of motion along $\bm{x}$ (blue squares), $\bm{y}$ (gray triangles) and $\bm{z}$ (red circles). In the simulation, kick duration is kept constant at $\tau=0.6\, t_0$ for $\bf z$ and $\tau = 3 \, t_0$ for $\bf x$, with increasing values of displacement $\delta$. The effective temperatures along $\bm{x}$ and $\bm{y}$ are equal and almost totally decoupled from $\bm{z}$. Solid lines are the best quadratic fits to the data: $\Delta T_\mathrm{x} / T_0 = \Delta T_\mathrm{y} / T_0 = 0.48 \times \eta^2$ and $\Delta T_\mathrm{z} / T_0 = -0.006 \times \eta^2$ for a kick along $\bm{x}$ and  $\Delta T_x /T_0 = \Delta T_y /T_0 = -0.006 \, \eta^2$, $\Delta T /T_0 = 0.52 \, \eta^2$  for a kick along $\bm{z}$. The dashed line in (a) is given by equation (\ref{Eq:pred}), assuming zero cross-thermalization between $\bm{z}$ and $\bm{x}$.}
	\label{fig:simukick}	
\end{figure}

More quantitatively, the relation between the center-of-mass momentum (resp. center-of-mass position for Weyl particles) after the kick and the effective temperature in the steady state can be approximated by a quadratic relation $\Delta T_{x,y,z} / T_0 = \alpha_{x,y,z} \eta ^2$, where the heating coefficients $\alpha_{i}$ depend on the kick direction, $\delta$ and $\tau$. For short excitation times, $\alpha_i$ are nearly independent of $\tau$. Their Their explicit dependence on $\delta$ is depicted in Fig. 4.

For kicks along $\bm{z}$, $\alpha_z$ does not vary significantly with the trap displacement for the experimentally relevant choice $\delta > 1$, in which case $\alpha_z=\alpha_0 = 0.5$. The value of $\tau$ essentially sets the strongest achievable kick $\eta$ and we take $\tau = 0.6 \, t_0$ in the simulation to cover the experimental range of excitations. The heating coefficient $\alpha_0 = 0.5$ is in agreement within error bars with the experimental result $\alpha_z = 0.63 (7)_\mathrm{stat} (20)_\mathrm{syst}$. The decoupling of the $\bf{x}$ direction appears more pronounced in the simulation than in the experiment with $ \Delta T_x /T_0 = -0.006 \, \eta^2$, to be compared  to the experimental value $\Delta T_\mathrm{x} /T_0 = -0.14 (5)_\mathrm{stat}(8)_\mathrm{syst} \times \eta^2$, a difference we attribute to imperfections of the magnetic excitation procedure.

For kicks along $\bm{x}$, $\alpha_x$ strongly varies with the kick amplitude $\delta$ (blue points in Fig. 4) and therefore a quantitative comparison with experiment requires a detailed modeling of the shape of the transient excitation currents, which is difficult. Nevertheless, fitting the duration $\tau = 3 t_0$ leads to $\Delta T_\mathrm{x} / T_0 = 0.48 \times \eta^2$ (to be compared to $\Delta T_\mathrm{x} / T_0 = 0.52 (5)_\mathrm{stat} (20)_\mathrm{syst} \times \eta^2$) and $\Delta T_\mathrm{z} / T_0 = -0.006 \times \eta^2$ (to be compared to $\Delta T_\mathrm{z} / T_0 = 0.10 (4)_\mathrm{stat} (5)_\mathrm{syst} \times \eta^2$). The chosen duration $3 \, t_0$ is consistent with the decay time of the eddy currents in our chamber ($\sim$ 3 ms).

\begin{figure}
\begin{center}
\includegraphics[width=0.80\columnwidth]{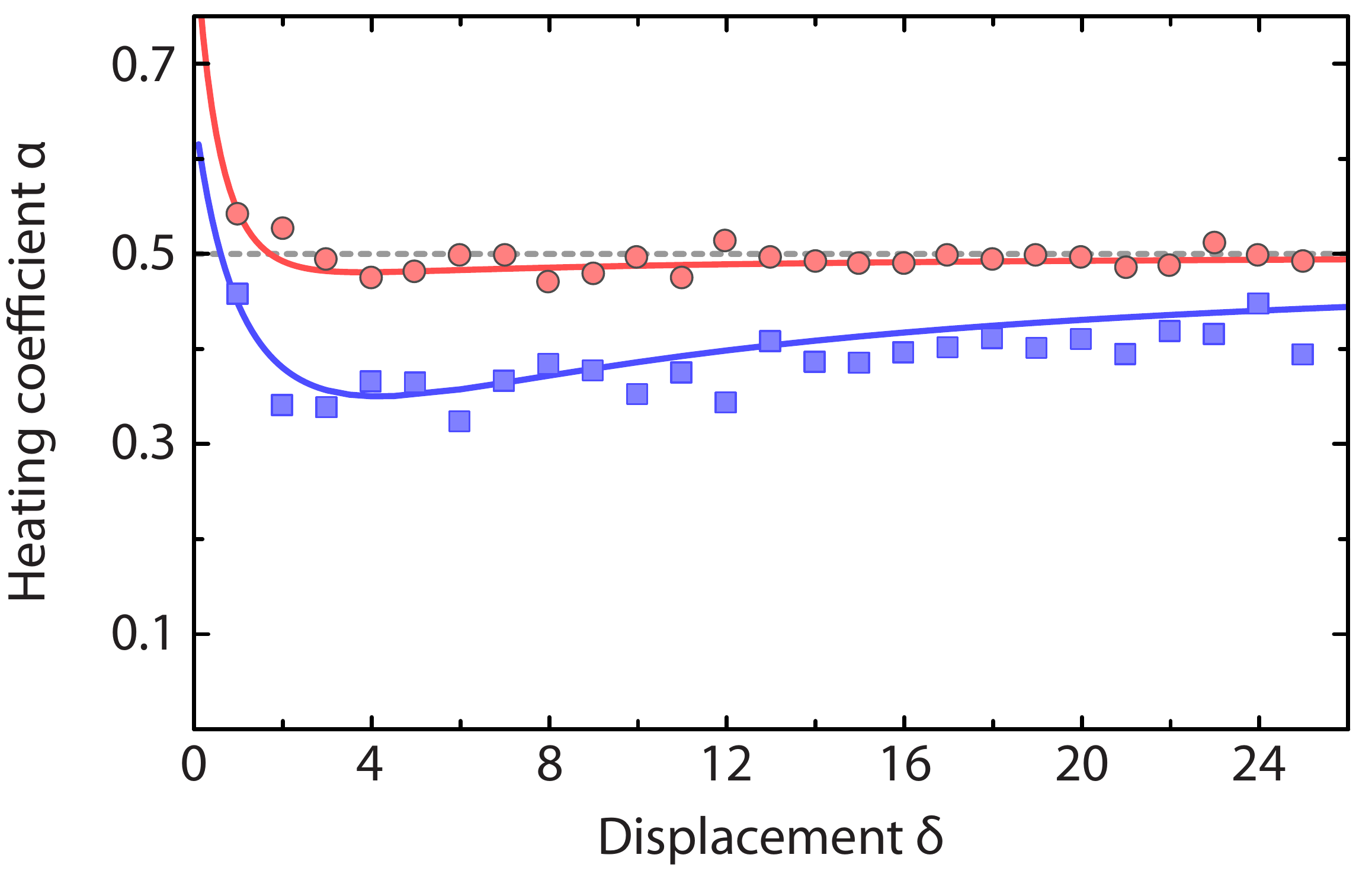}
\caption{Heating coefficient $\alpha$ along the kick direction versus Weyl point displacement $\delta$ in momentum space (resp. trap center displacement in position space) for kicks along $\bm x$ (blue) and $\bm z$ (red). $\alpha$ is defined as $ \alpha=\Delta T / (T_0 \eta ^2)$, relating excess temperature to kick strength $\eta$ (see text). For kicks along $\bf z$, $\alpha_z\sim0.5$ and is almost constant. On the contrary, for kicks along $\bf x$, $\alpha_x$ shows a strong dependence on displacement $\delta$. Solid lines are derived from equations (\ref{eq:DE2})-(\ref{eq:dE}) and (\ref{eq:Tz})-(\ref{eq:Txy}). Filled symbols are results from numerical simulations.}
\label{fig:DeltaE}
\end{center}
\end{figure}

\subsection{A simple model}
The heating of Weyl particles along the excitation direction can be understood from the the constraints imposed on the dynamics by energy conservation and virial theorem $E_{\rm kin}=2 E_{\rm pot}$. Here $E_{\rm kin}=\langle Pc\rangle$ and $E_{\rm pot}=\sum_i k_i \langle X_i^2\rangle/2$ are respectively the kinetic and potential energy of the Weyl particles, and the relation can be derived from its equivalent for massive particles in a linear trap. However, these two conditions are not sufficient to predict the final thermodynamic properties of the system. We therefore make two additional assumptions motivated by the results of the experiment and the simulations. (i) Heating occurs predominantly along the kick direction and (ii) whatever the kick's orientation may be, the final temperatures along the $x$- and $y$-directions are equal by symmetry. Under these conditions one finds for the final temperatures,
\begin{eqnarray}
z-{\rm kick:}&\Delta T_{\mathrm{x}}=\Delta T_{\mathrm{y}} \ll \Delta T_z,  &\Delta T_{\mathrm{z}}\simeq\frac{2\Delta E}{3k_B}, \label{eq:Tz}\\
x-{\rm kick:}&\Delta T_{\mathrm{x}}=\Delta T_{\mathrm{y}}\simeq\frac{\Delta E}{3k_B}, &\Delta T_z \ll \Delta T_{x,y}, \label{eq:Txy}
\end{eqnarray}
where $\Delta E$ is the energy transferred to the cloud through the excitation. Our numerical simulations satisfy (\ref{eq:Tz}) - (\ref{eq:Txy}) for the redistribution of the imparted energy.

In order to relate $\Delta E$ to the experimental kick strength $\eta$, we describe the dynamics of the cloud during the excitation through Liouville's equation for the phase-space density $f(\bm R,\bm P,t)$,
\begin{eqnarray}
	\partial_t f\left( {\bm R},{\bm P} ; t\right) 	&=&  -	\mathcal{L}f\left( {\bm R},{\bm P} ; t\right). \label{eq:Liouville}
\end{eqnarray}
The Liouville operator is defined as $\mathcal{L}=\partial_{\bm P}H_+^{\rm exc}.\partial_{\bm{ R}} - \partial_{\bm R}H_+^{\rm exc}.\partial_{\bm{P}}$ with $H_+^{\rm exc}=H_+(\bm R,\bm P-\bm \delta)$ being the shifted Weyl-point Hamiltonian. The formal solution to this equation is $f(\bm R,\bm P,\tau)=\exp(-\tau \mathcal{L})[f_0]$; for small excitation times $\tau$, we can Taylor-expand this expression an	d obtain
\be
	\langle\bm R\rangle = \tau  \int d^3\bm{r} d^3\bm{p} \, f_0\,(\bm{V}_{\rm exc}-\bm{V}),\label{eq:DE2}
\ee
\be
	\Delta E  =  \frac{\tau^{2}}{2}\int d^3\bm{r} d^3\bm{p} \, f_{0}\sum_i k_i\left({\bm V}_{\rm exc}-\bm{V}\right)_i^2,   \label{eq:dE}
\ee
where $\bm{V} = \partial_{\bm{P}} H_+$ is the velocity and $\bm V_{\rm exc}= \partial_{\bm P}H_+^{\rm exc}$.The relative scalings of $\Delta E$ and $\langle \bm R \rangle$ with $\tau$ confirm that $\alpha\propto \Delta E/\langle \bm R\rangle^2$ does not depend on the excitation duration in the short time limit. The values of $\alpha$ corresponding to equations (\ref{eq:DE2})-(\ref{eq:dE}) and (\ref{eq:Tz})-(\ref{eq:Txy}) are presented as solid lines in Fig. 4 and confirm the validity of the simulations.

For kicks along the $\bf z$ direction, we estimate the value of $\alpha_z$ by considering large displacements $\delta$, leading to
\begin{equation}
\Delta E = \frac{3}{4} E_0 \eta ^2.
\end{equation}
Interestingly, the energy gain is in fact larger than the value $E_0 \eta^2 / 2$ associated with the center of mass shift, because the cloud also expands in momentum space during the excitation whereby it gains additional kinetic energy. Inserting these asymptotic developments in equations (\ref{eq:Tz}) and (\ref{eq:Txy}), we finally obtain for $\bm{z}$ kicks the relative temperature increase along the excitation direction
\begin{equation}
\frac{\Delta T}{T_0}=\frac{\eta^2}{2}, \label{Eq:pred}
\end{equation}
corresponding to $\alpha_z = 0.5$, as discussed above and found in fair agreement with the experimental value.

\section{Conclusion}
Contrary to massive particles, Weyl fermions do not obey Kohn's theorem \cite{Kohn1962cyclotron} stating that the center of mass of an ensemble of non-relativistic massive particles oscillates  in a 3D harmonic potential without dephasing at frequencies $\sqrt{k_i/m}$. Instead, after an excitation, Weyl fermions move at constant speed even in a quadratic potential. Dephasing of the single-particle trajectories gives rise to damping of the center of mass motion and to an anisotropic spread of the position distribution, corresponding to an effective heating. In the symmetry plane, the steady-state distribution is almost decoupled from the strong axis but reach the same effective temperature along both directions regardless of the kick orientation, displaying a quasi-thermalization.

It should also be pointed out the anisotropic heating is not specific to our choice of spring constants for harmonic trap (\ref{Eq:H0Weyl}), which are in turn constrained by the mapping from the quadrupole potential. Additional simulations have shown that the same behavior is observed for arbitrary anisotropic potentials $V(\mathbf{r}) = (k_0 x^2 + k_0  y^2 + k_z z^2)/2$. Even in a fully isotropic situation $k_z = k_0$, the two unexcited directions are partially decoupled from the excited one and reach the same final temperatures, as the kick orientation breaks the overall symmetry.

It is crucial to note that in our experiments the energy transfer from the center-of-mass to the internal energy of the distribution does not depend on interactions between particles. It is  solely due to the complexity of the single particle trajectories in phase-space \cite{Surkov1994}, which originates from the non-harmonicity and non-separability of the underlying Hamiltonian (\ref{eq:H0'}). This absence of collisions is responsible for the non-thermal nature of the final distribution. Indeed, according to Thermodynamics' Second Law, Boltzmann's distribution maximizes the entropy of the system for a given energy. In our experiment, we start with  a thermal cloud characterized by a total energy $E$ and equilibrium entropy $S(E)$. A perfect momentum kick delivers an additional energy $\Delta E$ per particle, but does so without increasing the system's entropy. The latter is then conserved throughout the ensuing evolution because the ensemble remains collisionless. The quasi-equilibrium state thus exhibits a larger energy $E+\Delta E$ for the same entropy $S$, in contradiction to the usual entropy growth expected for a collisional system. The absence of real thermalization is then revealed by the anisotropic temperatures measured in the long time limit.
Weyl particles in a harmonic trap therefore provide an intriguing case of quasi-thermalization, midway between massive particles that do not equilibrate and collisional systems that reach a real Boltzmann thermal equilibrium (like in \cite{Davis1995}). As shown in \cite{chomaz2005generalized}, this situation can nevertheless be described within the framework of generalized Gibbs-ensembles as integrable systems in which a large number of constants of motion - here, the single-particle hamiltonian of individual atoms - prevents true thermalization \cite{rigol2007relaxation}.

Finally, the canonical mapping presented here  is not limited to the simulation of Weyl particles, but can address a broader range of problems. For instance, in a Ioffe-Pritchard trap a bias field gives rise to a non-zero magnetic field at the trap center and the overall field is of the form $B=\sqrt{B_0^2+b^2\sum_i \alpha_i^2 x_i^2}$. In this case, the analog system would be described by the relativistic kinetic energy $E=\sqrt{m^2c^4+p^2c^2}$ where the mass can be tuned through $B_0$. Another interesting situation arises in a hybrid trap consisting of the superposition of an optical dipole trap and a 2D magnetic quadrupole trap, where the Hamiltonian takes the form $h = \frac{p^2}{2m} + \frac{m\omega^2}{2}(x^2+y^2)+\frac{m\omega_z ^2 z^2}{2} - \mu_\mathrm{B}b(\sigma_x x - \sigma_y y)$. Applying our mapping to the variables $(x,y,p_x,p_y)$ leads to the equivalent Hamiltonian $H = \frac{P^2}{2m} + \frac{m\omega^2}{2}(X^2+Y^2)+\frac{m\omega_z ^2 Z^2}{2} - \frac{\mu_\mathrm{B}b}{m\omega}(\sigma_x P_x + \sigma_y P_y)$, which turns out to describe a 2D spin-orbit coupled particle  \cite{Koller2015}. Finally, in the same trap, it is also possible to engineer a Rashba coupling by taking $X=p_y/m\omega,  \, P_x=-m\omega y, \, Y = p_x/m\omega, \, P_y = -m\omega_x $.

\acknowledgments
The authors would like to thank  J. Dalibard and F. Gerbier for stimulating discussions. We are grateful to E. Demler for pointing out to us the analogy to massless relativistic particles. O. Goulko  acknowledges support from the NSF under the Grant No. PHY-1314735. C. Salomon expresses his gratitude to the A. von Humboldt foundation and to Prof. I. Bloch and T. W. H\"ansch for their kind hospitality at LMU and MPQ. This work was supported by R\'egion \^Ile de France (Dim nanoK/IFRAF), Institut de France (Louis D. prize) and the European Union (ERC grant ThermoDynaMix).

\end{document}